\documentclass{iau}
\usepackage{graphicx}

\title[JD 11.~~Exoplanet Dynamics] 
{Dynamical Constraints on Exoplanets}

\author[Horner, Wittenmyer, Tinney, Hinse \& Marshall]   
{Jonti Horner$^1$,
Robert A. Wittenmyer$^1$,
Chris Tinney$^1$, \\
Tobias C. Hinse$^{2}$,
 \and Jonathan P. Marshall$^3$}

\affiliation{$^1$School of Physics, UNSW, Sydney, NSW 2052, Australia. \\[\affilskip]
$^2$Korea Astronomy \& Space Science Institute, 305-348 Daejeon, Republic of Korea \\
$^3$Departamento de F\'{i}sica Te\'{o}rica, Facultad de Ciencias, Universidad Aut\'{o}noma de Madrid, Cantoblanco, 28049 Madrid, Spain}

\pubyear{2013}
\volume{299}  
\pagerange{ -- }
\jname{Exploring the Formation and Evolution of Planetary Systems}
\editors{B. Matthews \& J. Graham, eds.}
\begin{document}

\maketitle

\begin{abstract}
Dynamical studies of new exoplanet systems are a critical component of the discovery and characterisation process. Such studies can provide firmer constraints on the parameters of the newly discovered planets, and may even reveal that the proposed planets do not stand up to dynamical scrutiny. Here, we demonstrate how dynamical studies can assist the characterisation of such systems through two examples: QS Virginis and HD 73526.
\\
\\
\centering{\bf{To appear in the proceedings of IAU Symposium 299, accepted August 16, 2013}}

\end{abstract}

\vspace*{-0.6cm}
\section{Introduction}

In recent years, the number of known multiple exoplanet systems has risen dramatically. For those planets found by indirect means, the nature of the observations can cause problems in relating the behaviour of a given star to the orbits of the potential planets moving around it. The fitting of radial velocity data, for example, is known to be prone to over-estimating the eccentricity of a given planet's orbit (e.g. \cite{Ecc1}, \cite{Ecc2}). Indeed, once further data is obtained, a system thought to host a single, highly eccentric planet may be found to instead host two planets moving on markedly different, low eccentricity orbits (e.g. Wittenmyer \textit{et al.}, 2012, 2013b). 

One way these uncertainties can be reduced is to consider the interaction between the proposed planets. For systems in which the calculated orbits represent the true state of the system, it is reasonable to assume that they should be dynamically stable on long timescales . On the other hand, if the proposed orbits for the planets prove dynamically unstable on timescales of thousands of years, then this is a strong indication that the true orbits of the proposed planets differ markedly from those put forward solely on the basis of the observational data. In this work, we present two examples of how dynamical studies can help to untangle the nature of recently proposed planetary systems. In section two, we briefly describe our methods, before presenting and discussing our results for two exemplar systems in section three. 

\vspace*{-0.5cm}
\section{Methods}
To assess the stability of a given planetary system, we follow a well established methodology (e.g. \cite{Dyn1, Dyn2}). We use the Hybrid integrator within the dynamics package \textsc{Mercury} (\cite{Merc}) to follow the evolution the system. In those simulations, we hold the initial orbit of the best categorised planet fixed, and systematically vary the initial orbital parameters of the other planet. The orbital evolution of the two planets is followed for a period of 100 Myr, or until one or other of the planets is removed from the system by collision or ejection. This allows us to create dynamical maps of the system, revealing the degree to which the orbits of the planets are dynamically stable. For more details, we refer the interested reader to our earlier work (e.g. Horner \textit{et al.}, 2011, 2012, \cite{Ecc4}).

\vspace*{-0.5cm}
\section{Two Exemplar Systems}
In 2011, two unseen companions were proposed to orbit the cataclysmic variable QS Virginis (\cite{QSVir1}), based on eclipse timing variations. The companions, of 9 and 57 Jupiter masses, move on crossing orbits, with semimajor axes of $6.031$ and $7.043$~AU and eccentricities of $0.62$ and $0.92$. The system's stability, as a function of the orbit of QS Vir c, can be seen in the left-hand panel of Figure 1. The mean lifetime is just a few hundred years across the entire range of plausible orbits for QS Vir (AB) c. The companions are clearly dynamically unfeasible (\cite{QSVir}).

HD 73526~c was proposed on the basis of radial velocity data obtained by the Anglo-Australian Planet Search (\cite{73526}). The planets are located close to, or within, mutual 2:1 resonance. The results of our dynamical study are shown in the right hand panel of Figure 1. In this case, resonant solutions exhibit dynamical stability on timescales of millions of years. A large sea of unstable solutions surrounds the narrow region of resonant stability. Our dynamical study therefore enables us to place tighter constraints on the orbit of HD 73526~c than were possible based solely on the observations. 

\begin{figure}
\centering
\begin{tabular}{cc}
\includegraphics[width=3.0in, height=2.1in]{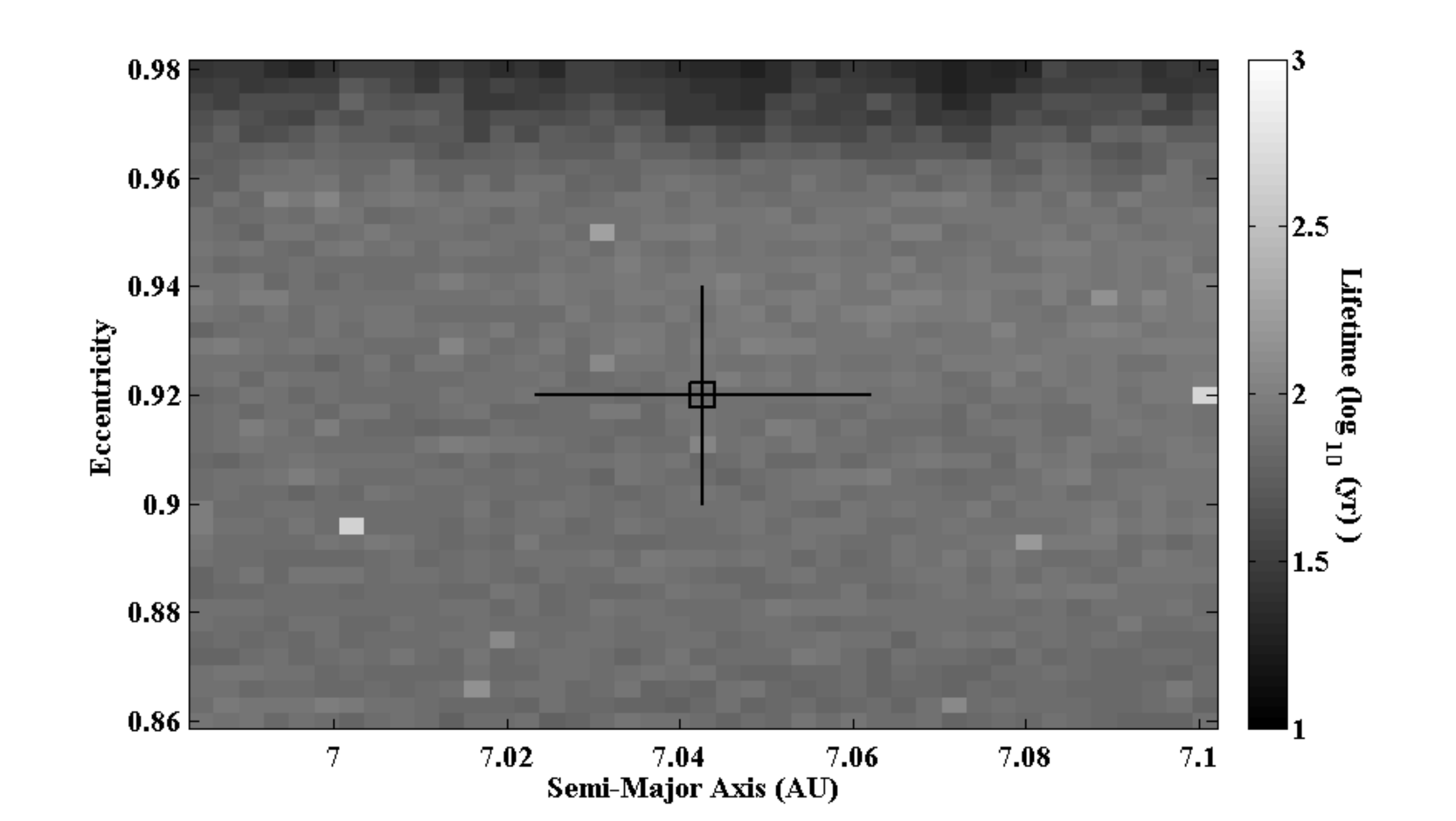} & 
\includegraphics[width=3.0in, height=2.1in]{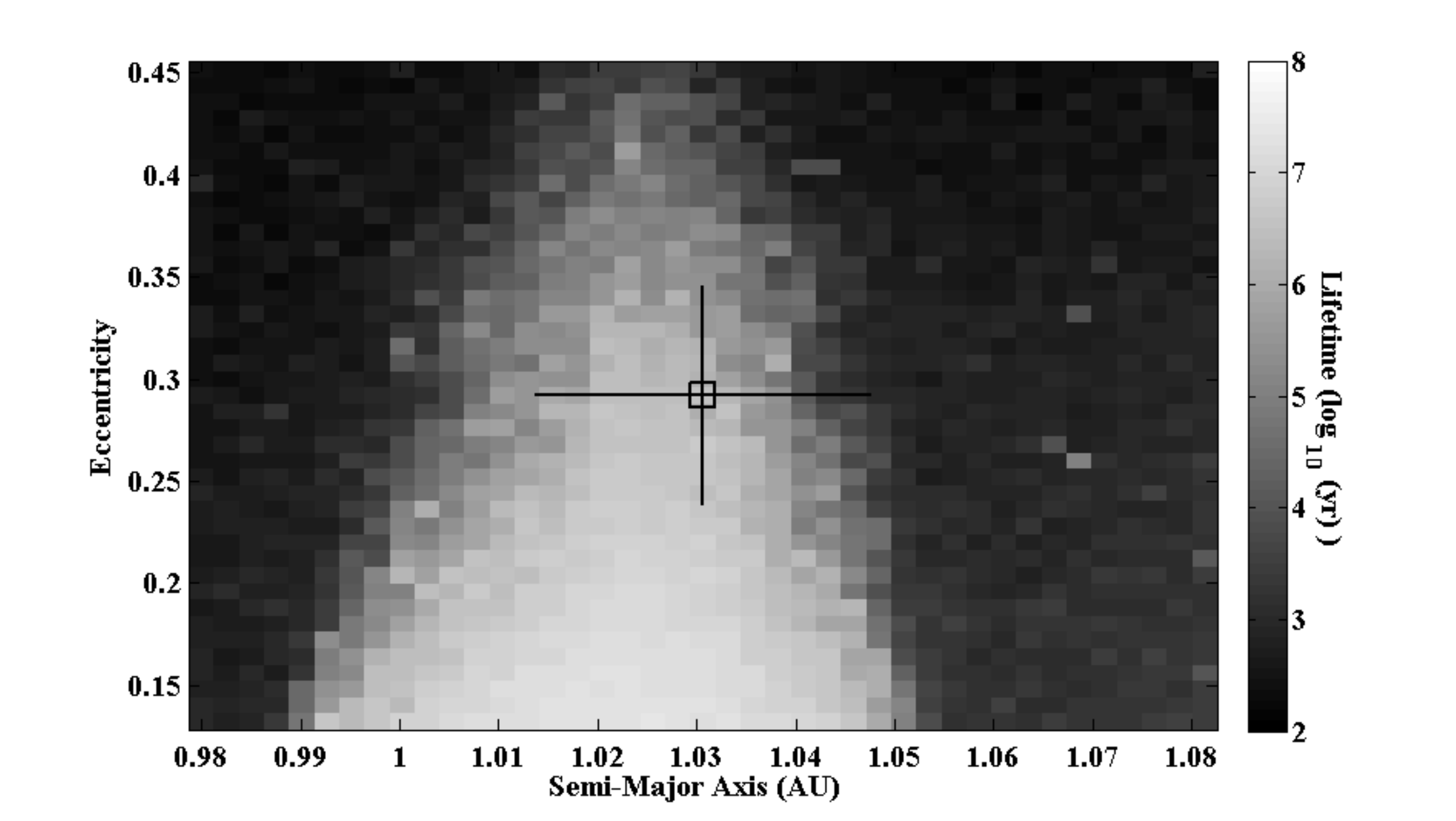} \\
\end{tabular}
\caption{The stability of QS~Vir (left) and HD~73526 (right). The best-fit orbit for the outer planet is denoted by the hollow box, with 1-$\sigma$ errors shown by the crosshairs.}
\end{figure}

\end{document}